\documentclass[aps,prl,reprint,showpacs,floatfix,superscriptaddress,citeautoscript]{revtex4-1}
\usepackage{graphicx}
\usepackage{bm,amsmath,amssymb,mathrsfs,dcolumn}
\usepackage{epsfig}
\usepackage{hyperref}
\usepackage[usenames]{color} 
\hypersetup{pdfborder=0 0 0,colorlinks=true,citecolor=blue,linkcolor=blue}
\usepackage[final]{pdfpages}

\begin{document}
\author{Yi Liu}
\affiliation{Faculty of Science and Technology and MESA$^+$ Institute for Nanotechnology, University of Twente, P.O. Box 217, 7500 AE Enschede, The Netherlands}
\author{Zhe Yuan}
\affiliation{Faculty of Science and Technology and MESA$^+$ Institute for Nanotechnology, University of Twente, P.O. Box 217, 7500 AE Enschede, The Netherlands}
\affiliation{Institut f{\"u}r Physik, Johannes Gutenberg--Universit{\"a}t Mainz, Staudingerweg 7, 55128 Mainz, Germany}
\author{R. J. H. Wesselink}
\author{Anton A. Starikov}
\affiliation{Faculty of Science and Technology and MESA$^+$ Institute for Nanotechnology, University of Twente, P.O. Box 217, 7500 AE Enschede, The Netherlands}
\author{Mark van Schilfgaarde}
\affiliation{Department of Physics, Kings College London, London WC2R 2LS, United Kingdom}
\author{Paul J. Kelly}
\affiliation{Faculty of Science and Technology and MESA$^+$ Institute for Nanotechnology, University of Twente, P.O. Box 217, 7500 AE Enschede, The Netherlands}

\title{Direct Method for Calculating Temperature-Dependent Transport Properties} 
\date{\today}
\begin{abstract}
We show how temperature-induced disorder can be combined in a direct way with first-principles scattering theory to study diffusive transport in real materials. Excellent (good) agreement with experiment is found for the resistivity of Cu, Pd, Pt (and Fe) when lattice (and spin) disorder are calculated from first principles. For Fe, the agreement with experiment is limited by how well the magnetization (of itinerant ferromagnets) can be calculated as a function of temperature. By introducing a simple Debye-like model of spin disorder parameterized to reproduce the experimental magnetization, the temperature dependence of the average resistivity, the anisotropic magnetoresistance and the spin polarization of a Ni$_{80}$Fe$_{20}$ alloy are calculated and found to be in good agreement with existing data. Extension of the method to complex, inhomogeneous materials as well as to the calculation of other finite-temperature physical properties within the adiabatic approximation is straightforward.
\end{abstract}
\maketitle

{\color{red}\it Introduction.---}Measuring the temperature dependence of electrical transport is one of the most important and common experimental probes of condensed matter. Although a great deal of what determines the temperature dependence is understood qualitatively \cite{Ziman:60}, there has been virtually no progress in translating this understanding into quantitative, material-specific studies in the past twenty years because of the complexity of the theoretical formalisms \cite{Allen:96, Savrasov:prb96b}; the lowest order variational approximation (LOVA) that is the basis for the successful description of the temperature-dependent electrical and thermal resistivities of a number of elemental metals \cite{Savrasov:prb96b} has to the best of our knowledge not been applied to more complex materials. In particular, it has not been extended to the study of magnetic materials. The need to be able to do so is pressing because current studies of magnetization switching involve large threshold current densities that are accompanied by substantial Joule heating \cite{Stamps:jpcm14}.

\begin{figure}[b]
\includegraphics[width=0.95\columnwidth]{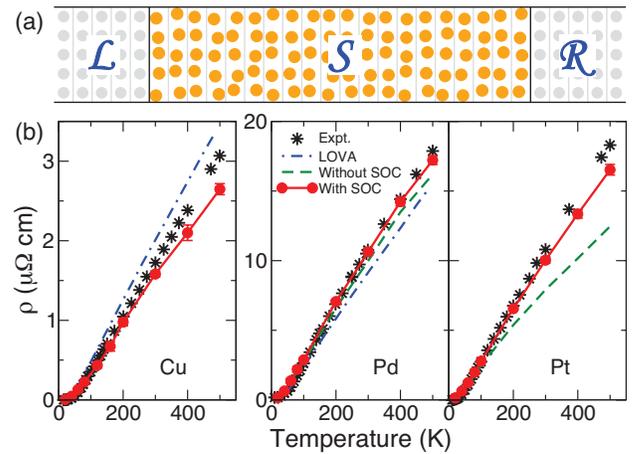}
\caption{(a) Illustration of the scattering geometry used to calculate transport properties. By populating first-principles phonon modes, we generate correlated lattice disorder in a scattering region ($\mathcal S$) that is connected to semi-infinite, crystalline left ($\mathcal L$) and right ($\mathcal R$) leads. (b) Temperature-dependent electrical resistivities calculated for Cu, Pd and Pt. The green dashed lines for Pd and Pt are results obtained without SOC. Experimental data (black stars) \cite{Bass:LB82,AIPhbk72} and the results of LOVA calculations (blue dash-dotted lines) \cite{Savrasov:prb96b} are shown for comparison.}\label{fig:1}
\end{figure}

Inspired by the success of the ``direct'' {\it ab initio} molecular dynamics approach to studying structural and electronic properties of matter at finite temperatures   introduced by Car and Parrinello \cite{Car:prl85}, we have developed a direct approach to calculate finite-temperature transport properties within the adiabatic approximation. For nonmagnetic (NM) materials, we generate ``snapshots'' of a thermally disordered  solid \footnote{Using {\it ab initio} lattice \cite{Baroni:prl87,Frank:prl95,Kresse:epl95} or molecular \cite{Car:prl85} dynamics, or using information from experiment.} and use first-principles scattering theory to determine the scattering matrix \cite{Xia:prb01,Xia:prb06} and related properties \cite{Brataas:prp06}, Fig.~\ref{fig:1}(a). The results of this two-stage procedure are illustrated by comparing the calculated and experimentally measured temperature-dependent resistivities of the NM metals Cu, Pd and Pt in Fig.~\ref{fig:1}(b). The purpose of this Letter is to underpin and extend these extremely promising results by including spin-orbit coupling (SOC) \cite{Starikov:prl10,Liu:prb11,Yuan:prl12,Liu:prl14,Yuan:prl14} to determine the temperature dependence of the spin-flip diffusion lengths $l_{\rm sf}$ for Pd and Pt, and for ferromagnetic (FM) materials, to include spin-disorder \cite{Liu:prb11}.

{\color{red}\it NM metals.---}To describe thermally induced lattice disorder in NM metals, we first use density functional theory to calculate the dynamical matrix of the bulk metal \cite{Baroni:prl87,Frank:prl95,Kresse:epl95,SM1}. The eigenvalues and eigenvectors that result from diagonalizing this matrix correspond to phonon energies and vibrational polarizations, respectively. These are used to construct snapshots of correlated thermal lattice disorder by superposing all allowed modes in the scattering region populated at a chosen temperature \cite{SM1}. The disordered region is connected to ideal leads to perform scattering calculations using the Landauer-B{\"u}ttiker formalism implemented with tight-binding muffin-tin orbitals \cite{Xia:prb06} with atomic sphere potentials displaced rigidly with the atoms; see Fig.~\ref{fig:1}(a). For the chosen temperature, the conductance is calculated for a number of configurations of disorder and the resistivity extracted by varying the length of the scattering region as in Refs.\cite{Starikov:prl10,Liu:prb11}, whereby the influence of the leads is eliminated. This static lattice disorder scheme for electronic transport is based upon the Born-Oppenheimer approximation, which is justified because typical momentum relaxation times for conduction electrons in metals (10$^{-14}$--10$^{-15}$~s) are so much shorter than the time scale of atomic vibrations (10$^{-12}$~s).

Without introducing any adjustable parameters, the calculated resistivities of Cu, Pd and Pt, plotted as red circles in Fig.~\ref{fig:1}(b), are seen to be in very good agreement with experiment (black stars) \cite{Bass:LB82,AIPhbk72} indicating that our computational scheme captures the main physics of the electron-phonon interaction in NM metals. Although SOC has very little effect on the calculated phonon spectra, including it in the transport calculations for Pd and Pt increases the resistivity and improves the agreement with experiment. In particular, above room temperature the calculated resistivity of Pt with SOC is 30\% higher  highlighting its importance for 5$d$ transition metals. SOC lifts the degeneracy of energy bands and changes the shape of the Fermi surface. By allowing spin-flipping, it modifies the phase space of final states that can be reached by electron-phonon scattering. The SOC strength for $d$ electrons at the Fermi energy in Pd (18 mRy) is only about one third of that in Pt (55 mRy) so that its effect on the resistivity of Pd is much weaker. In addition to the SOC parameter for Cu being much smaller, the electronic states at its Fermi level have mainly $s$ character ($l=0$); including SOC has negligible effect on the calculated resistivity. For comparison, we also plot in Fig.~\ref{fig:1}(b) the results from calculations that used LOVA \cite{Ziman:60,Pinski:prb81} to solve the Boltzmann equation \cite{Savrasov:prb96b} (blue dash-dotted lines). For the past twenty years, this has been the state-of-the-art.

\begin{figure}[t]
\includegraphics[width=0.9\columnwidth]{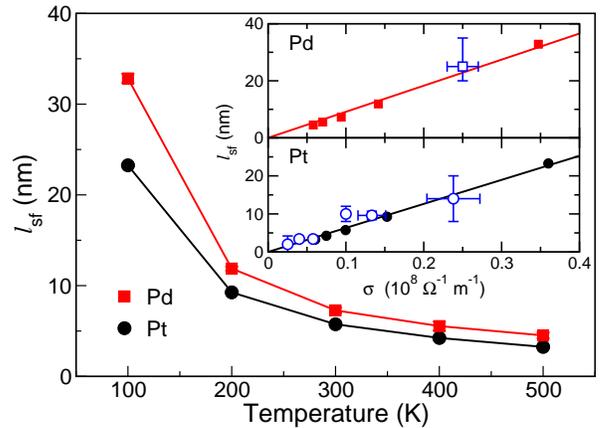}
\caption{Calculated spin-flip diffusion length of Pd and Pt as a function of temperature. Inset: spin-flip diffusion length of Pd and Pt replotted as a function of conductivity (solid symbols). The solid lines illustrate the linear dependence. Experimental values that are either not sensitive to interface spin flipping \cite{Niimi:prl13,Isasa:prb15} or take it into account \cite{Kurt:apl02,Nguyen:jmmm14,Rojas-Sanchez:prl14} are plotted for comparison (empty blue symbols).}\label{fig:2}
\end{figure}

In NM metals like Pd and Pt, the spin-flip diffusion length $l_{\rm sf}$ characterizes the relaxation of a longitudinal nonequilibrium spin distribution and is an important material parameter that enters the description of many spin phenomena like the spin Hall effect, spin pumping, etc. \cite{Tserkovnyak:prb02b, Liu:arXiv11, Rojas-Sanchez:prl14, Nguyen:jmmm14, Liu:prl14}. Its temperature dependence is particularly interesting since many experiments are performed at room temperature; for current-switching experiments, Joule heating can increase the temperature of the material significantly so electron-phonon scattering is unavoidable. By injecting a fully spin-polarized current into a disordered NM metal, we are able to obtain $l_{\rm sf}$ using an exponential fit to the calculated spin-resolved conductance as a function of the length of the scattering region \cite{Starikov:prl10,Liu:prl14}. The values of $l_{\rm sf}$ we calculate for Pd and Pt are shown in Fig.~\ref{fig:2}. Both decrease with increasing temperature as $1/T$. Pd has a larger $l_{\rm sf}$ than Pt at the same temperature because its weaker SOC leads to a smaller probability of disorder-induced spin-flip scattering. 

Unlike bulk resistivities that are generally well-documented, the values of $l_{\rm sf}$ extracted from experiment for Pt and Pd exhibit a spread of more than an order of magnitude \cite{Isasa:prb15,Boone:jap13,Kurt:apl02}. One reason for this spread is the neglect of interface spin-flip scattering when interpreting spin-pumping experiments. This leads to a severe underestimation of $l_{\rm sf}$, especially for Pt \cite{Nguyen:jmmm14,Rojas-Sanchez:prl14,Liu:prl14}. Another reason is the variable purity of experimental samples as evidenced by low-temperature resistivities that differ substantially \cite{Kurt:apl02,Niimi:prl13,Nguyen:jmmm14}. Indeed, it has already been pointed out that there is no especially good reason to expect $l_{\rm sf}$ to depend only on temperature-induced disorder \cite{Nguyen:jmmm14}. Nguyen {\it et al.} present experimental evidence for a linear dependence of $l_{\rm sf}$ on the independently measured conductivity $\sigma=1/\rho$ \cite{Nguyen:jmmm14} in agreement with a relationship due to Elliott \cite{Elliott:pr54}. We replot $l_{\rm sf}$ as a function of $\sigma$ in the inset to Fig.~\ref{fig:2} and find a perfectly linear relation for both Pd and Pt. For comparison, we also plot data extracted from experiments that are either not sensitive to interface spin flipping \cite{Niimi:prl13,Isasa:prb15} or take it into account \cite{Kurt:apl02,Nguyen:jmmm14,Rojas-Sanchez:prl14}. The good agreement between theory and experiment suggests that the Elliott-Yafet mechanism \cite{Elliott:pr54,Yafet:63} dominates the spin relaxation in NM metals like Pd and Pt.

\begin{figure}
\includegraphics[width=0.9\columnwidth]{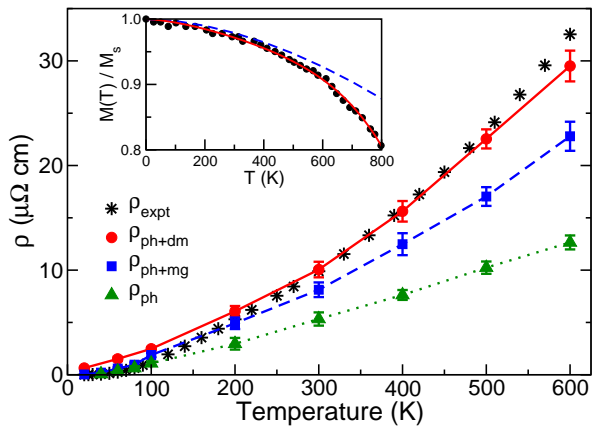}
\caption{Electrical resistivity for bcc Fe calculated as a function of the temperature. Green triangles: resistivity arising from phonon-induced lattice disorder; blue squares: resistivity in the presence of phonons and magnons that are both populated in the scattering region at a given temperature; red circles: resistivity with phonon lattice disorder and uncorrelated spin disorder that reproduces the experimental demagnetization curve (inset). The experimental values \cite{AIPhbk72} are plotted as black stars for comparison. Because of the smallness of the anisotropic magnetoresistance in Fe \cite{Isin:pr66,Tondra:jap93}, we only consider the case of magnetization parallel to the current direction. Inset: temperature-dependent magnetization of Fe from experiment (black dots) \cite{Crangle:prsla71} and obtained by populating magnons (blue dashed line). The red line interpolates the experimental values using a cubic spline method. }\label{fig:3}
\end{figure}

{\color{red}\it FM metals.---}In magnetic materials, temperature influences electrical transport by disrupting not only the translational periodicity of the lattice but also the magnetic ordering. Taking bcc FM Fe as an example, we carry out the same procedure as for NM metals to examine the resistivity resulting only from lattice disorder, $\rho_{\rm ph}$. The results, plotted in Fig.~\ref{fig:3} as green triangles, are seen to be much smaller than the measured values and scale approximately linearly with temperature above 50 K (as they do for NM metals), whereas the experimental data show a higher order dependence on temperature. This lack of agreement indicates the need to take other scattering mechanisms into account. 

The electrical resistivity arising from spin fluctuations has recently been studied using various models \cite{Wysocki:prb09, Kudrnovsky:prb12, Glasbrenner:prb14, Liu:prb11, Kovacik:prb14, Ebert:arXiv14}, none of which is fully {\it ab initio}. By analogy with the phonon description of lattice disorder just presented, we can introduce spin disorder without any adjustable parameters by superposing magnon modes that are populated as a function of temperature to generate snapshots of correlated spin disorder \cite{SM1}. In the absence of a magnon mass, this procedure can be justified by appealing to the low frequency of thermal spin fluctuations compared to typical electron relaxation times. We calculate the spin waves of bulk Fe using the ``frozen magnon method'' \cite{Halilov:epl97,*Halilov:prb98b} and then calculate the resistivity of Fe with both lattice and spin disorder, $\rho_{\rm ph+mg}$ in the same way that we calculated $\rho_{\rm ph}$. The results are shown in Fig.~\ref{fig:3} as blue squares. Below room temperature, the agreement with experiment is much improved. To the best of our knowledge this is the first attempt to include both lattice and spin disorder in a quantitative, parameter-free study of transport in a magnetic material.

In spite of the improvement, above room temperature $\rho_{\rm ph+mg}$ is lower than the experimental data. To understand the deviation, we calculate the magnetization corresponding to themally occupying the magnon modes \cite{Halilov:prb98b} and plot it in the inset to Fig.~\ref{fig:3} (blue dashed line) together with the experimentally measured thermal demagnetization curve (black dots) \cite{Crangle:prsla71}. It is seen that below room temperature, the calculated magnetization reproduces the measured values quite well but gradually deviates from the experimental data as the temperature increases. As the magnon mode occupancy increases with temperature, it gives rise to relatively large cone angles, especially for long wavelength modes. The frozen magnon method that is based upon a small cone angle assumption (analogous to the harmonic approximation) is no longer applicable. In addition, magnon modes start to interact with each other and deviate from pure bosonic character \cite{Halilov:prb98b}. 

To examine the effect of overestimating $M(T)$ more clearly, we switch to the empirical uncorrelated spin disorder scheme (disordered moment: dm) introduced in \cite{Liu:prb11} to reproduce the experimental magnetization at every temperature in addition to the correlated lattice disorder obtained by populating phonons. This results in a larger resistivity $\rho_{\rm ph+dm}$ (red circles in Fig.~\ref{fig:3}) that agrees remarkably well with experiment. Below room temperature, the lack of correlation in the spin disorder leads to a slightly higher resistivity than what is seen in experiment. We conclude that the underestimation of the fully {\it ab initio} scheme, $\rho_{\rm ph+mg}$, is attributable to the overestimation of the magnetization at higher temperatures. 

{\color{red}\it FM alloys.---}Within the adiabatic approximation, our ability to describe from first principles the temperature dependence of transport properties is limited by our ability to characterize temperature-dependent lattice and spin disorder. The limitations posed by the harmonic approximation of lattice dynamics could be circumvented by using {\it ab initio} molecular dynamics to generate suitable configurations of disorder. In spite of the many efforts made to improve the finite-temperature description of magnetism \cite{Gyorffy:jpf85,Kubler:jpcm06,Kormann:prb08}, this is still an essentially open problem. For FM materials with well characterized demagnetization curves, we can use the uncorrelated spin disorder scheme discussed above. We illustrate the effectiveness of this approach by calculating some temperature-dependent transport properties of the important FM alloy Ni$_{80}$Fe$_{20}$, Permalloy. For simplicity, we describe the (uncorrelated Gaussian) lattice disorder using a Debye model \cite{Liu:prb11,Ebert:prl11}. 

\begin{figure}[t]
\includegraphics[width=0.9\columnwidth]{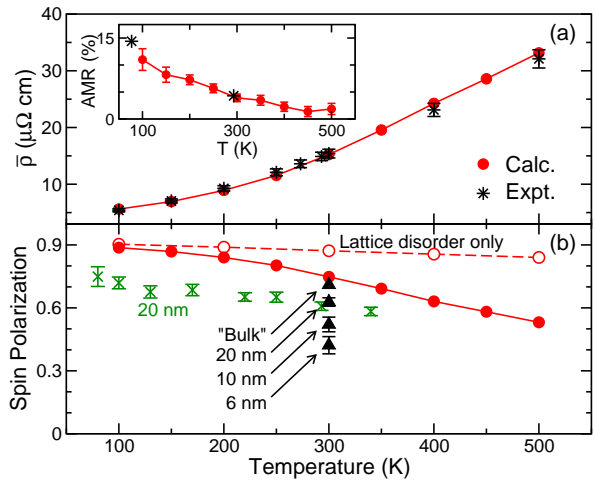}
\caption{Calculated temperature-dependent properties of the fcc Ni$_{80}$Fe$_{20}$ substitutional alloy, Permalloy. (a) Comparison of calculated and measured \cite{Ho:jpcrd83} average resistivities $\bar{\rho}(T)$. Inset: AMR as a function of the temperature. Experimental values \cite{McGuire:ieeem75} at 77~K and 293~K are plotted for comparison. (b) Spin polarization $P\equiv(j_{\uparrow}-j_{\downarrow})/(j_{\uparrow}+j_{\downarrow})$ for the magnetization parallel to the current direction. $P$  becomes only 1\% larger when the magnetization is perpendicular to the current direction. The empty (red) circles indicate the spin polarization with lattice disorder only. The (green) crosses are values of $P$ extracted from current-induced spin-wave Doppler shift experiments on a 20 nm thick film \cite{Zhu:prb10}. The black triangles are the results of measurements at room temperature for three film thicknesses and ``bulk'' is the value obtained by extrapolation to infinite thickness \cite{Haidar:prb13}.
}
\label{fig:4}
\end{figure}

Using the Debye temperature of 450~K extracted from experiment \cite{Tanji:jpsj71} and the experimentally measured magnetization \cite{Wakelin:ppsb53} to generate configurations of uncorrelated Gaussian lattice and spin disorder, we calculate the  temperature-dependent average resistivity $\bar{\rho}=(2\rho_{\perp}+\rho_{\parallel})/3$ for Permalloy shown in Fig.~\ref{fig:4}(a), where $\rho_{\perp}$ and $\rho_{\parallel}$ are the resistivities calculated with the magnetization perpendicular and parallel to the current direction, respectively. The anisotropic magnetoresistance (AMR) that leads to the spin rectification effect \cite{Bai:prl13} and is responsible for the negative domain-wall resistance in Permalloy \cite{Klaui:prl03,Yuan:prl12} is also calculated and plotted as a function of temperature in the inset to Fig.~\ref{fig:4}(a) as the AMR ratio $(\rho_{\parallel}-\rho_{\perp})/\bar{\rho}$. Both $\bar{\rho}$ and the AMR ratio are virtually indistinguishable from the best available experimental data. The AMR decreases from 11\% at 100~K to 4\% at room temperature. This reduction is consistent with a picture of SOC-induced scattering, whereby thermal lattice disorder gives rise to weaker anisotropic scattering than magnetic ``impurity'' (Fe in Ni) scattering \cite{Smit:phys51}.

The degree of spin polarization $P$ of a FM metal, called the spin asymmetry in the diffusion theory of spin transport \cite{Valet:prb93}, plays a very important role in many spintronics applications \cite{Marrows:ap05} but has been quite controversial \cite{Mazin:prl99}. It has been experimentally measured using different transport techniques, such as magnetoresistance \cite{Bass:jpcm07}, Andreev reflection \cite{deJong:prl95,Soulen:sc98}, current-induced domain-wall motion \cite{Curiale:prl12,Ueda:apl12} and spin-wave Doppler shift \cite{Vlaminck:sc08,Zhu:prb10}. However, the values reported for Permalloy do not agree with each other \cite{Soulen:sc98,Bass:jpcm07,Vlaminck:sc08,Zhu:prb10,Haidar:prb13}. Here we focus on the bulk value of $P\equiv(j_{\uparrow}-j_{\downarrow})/(j_{\uparrow}+j_{\downarrow})$ and its temperature dependence by examining the projected current densities $j_{\uparrow}$ parallel to and $j_{\downarrow}$ antiparallel to the quantization axis, respectively \cite{Wang:prb08}. Figure~\ref{fig:4}(b) shows the temperature dependence of $P$ calculated for bulk Permalloy (solid circles) for the magnetization parallel to the current direction; it increases by about $1\%$ when the current is perpendicular to the magnetization. As the temperature increases from 100 to 500~K, $P$ decreases monotonically from 0.9 to 0.53. The main contribution to the reduction arises from thermally induced spin fluctuations. If we artificially switch off spin disorder and only include lattice disorder, the calculated spin polarization (empty circles) is much larger, especially at high temperature. Experiment finds that $P$ depends not only on the temperature \cite{Zhu:prb10} but also on the thickness of the sample \cite{Haidar:prb13}, surface scattering apparently depolarizing the current. The value labelled ``bulk'' in Fig.~\ref{fig:4}(b) was measured at room temperature and obtained by extrapolating the results found for thin films \cite{Haidar:prb13}. The agreement with our bulk calculation is excellent \footnote{The  polarization presented here is that for bulk Permalloy (Py) while that calculated in Ref.~\cite{Starikov:prl10} was for a NM$|$Py interface. In general, these polarizations are not the same.}. 

{\color{red}\it Summary.---}We have presented a conceptually simple ``direct'' method for calculating temperature dependent transport properties based upon the adiabatic approximation that combines first-principles scattering theory with temperature-induced disorder modelled in large lateral supercells. The effectiveness of the procedure is illustrated by the very good agreement that we find between measured and calculated temperature-dependent resistivities of Cu, Pd, Pt, Fe and Permalloy. Because our scattering formalism includes SOC, its influence on transport in combination with temperature-induced disorder can be studied. We find good agreement with available experimental results for the spin-flip diffusion lengths of Pd and Pt and for the AMR and spin polarization of Permalloy. Our calculation of the intrinsic $l_{\rm sf}$ for bulk Pd and Pt should help to resolve the ongoing controversy about its value in different situations and provide a way to determine $l_{\rm sf}$ from measured resistivities. 

Temperature-induced disorder modeled in this way can be used to calculate other properties at finite temperatures such as densities of states, optical excitations, thermoelectric effects etc., as long as the adiabatic approximation is applicable. For example, in spin caloritronics \cite{Bauer:natm12}, a temperature gradient can be modeled as an inhomogeneous phonon and magnon occupation. Our methodology makes it possible to study how transport changes from ballistic to diffusive with temperature without building any assumption about the nature of the transport into the theoretical approach. This can be especially important for the complex, inhomogeneous layered structures that are essential for spintronics devices, where mean free paths are longer than the ``thickness'' of interfaces.

\begin{acknowledgments}
This work was financially supported by the ``Nederlandse Organisatie voor Wetenschappelijk Onderzoek'' (NWO) through the research programme of ``Stichting voor Fundamenteel Onderzoek der Materie'' (FOM) and the  supercomputer facilities of NWO ``Exacte Wetenschappen (Physical Sciences)''. It was also partly supported by the Royal Netherlands Academy of Arts and Sciences (KNAW). Z. Y. acknowledges the financial support of the Alexander von Humboldt Foundation.
\end{acknowledgments}

\pagebreak
\begin{widetext}
\clearpage
\begin{center}
\textbf{\large Supplemental Material for ``Direct Method for Calculating Temperature-Dependent Transport Properties''}

\vspace{3mm}
Yi Liu,$^1$ Zhe Yuan,$^{1,2}$ R. J. H. Wesselink,$^1$ Anton A. Starikov,$^1$ Mark van Schilfgaarde,$^{3}$ and Paul J. Kelly$^1$

\vspace{2mm}
$^1${\small\it Faculty of Science and Technology and MESA$^+$ Institute for Nanotechnology, \\University of Twente, P.O. Box 217, 7500 AE Enschede, The Netherlands}

$^2${\small\it Institut f{\"u}r Physik, Johannes Gutenberg--Universit{\"a}t Mainz, Staudingerweg 7, 55128 Mainz, Germany}

$^3${\small\it Department of Physics, Kings College London, London WC2R 2LS, United Kingdom}

\end{center}
\end{widetext}
\setcounter{equation}{0}
\setcounter{figure}{0}
\setcounter{table}{0}
\setcounter{page}{1}
\makeatletter
\renewcommand{\theequation}{S\arabic{equation}}
\renewcommand{\thefigure}{S\arabic{figure}}
\renewcommand{\thetable}{S\Roman{figure}}
\renewcommand{\bibnumfmt}[1]{[S#1]}
\renewcommand{\citenumfont}[1]{S#1}

When generating thermal lattice (and spin) disorder in the scattering region, the atomic (spin) displacments should have the periodicity of the lateral supercells. This can be done by calculating all phonon (magnon) modes with wavevectors $\mathbf q$ that correspond to reciprocal lattice vectors of the real space vectors that describe the supercell. 

\section{I. Phonon calculations}\label{sec:ph}
To calculate thermal lattice disorder, we use the ``force constant approach'' \cite{Kresse:epl95s,vanGelderen:prb03s}. In this method, a (quite unrelated) supercell of the material making up the scattering region is constructed, the central atom is displaced by a small amount ${\bm \delta}$ and the forces induced on all of the atoms in the supercell by this displacement are calculated. As long as it is sufficiently small, the forces should be linear in ${\bm \delta}$ allowing them to be differentiated numerically to form second derivatives of the energy, the force constant matrix. By constructing Bloch sums of the force constant matrix for arbitrary wave vector $\mathbf q$, we obtain the dynamical matrix. Because of the strong screening in metals, the force field induced by displacing a central atom is short ranged. Using elements of the force constant matrix calculated with a $5\times5\times5$ supercell yields well converged phonon dispersion relations.

\begin{figure}[b]
\includegraphics[width=1\columnwidth]{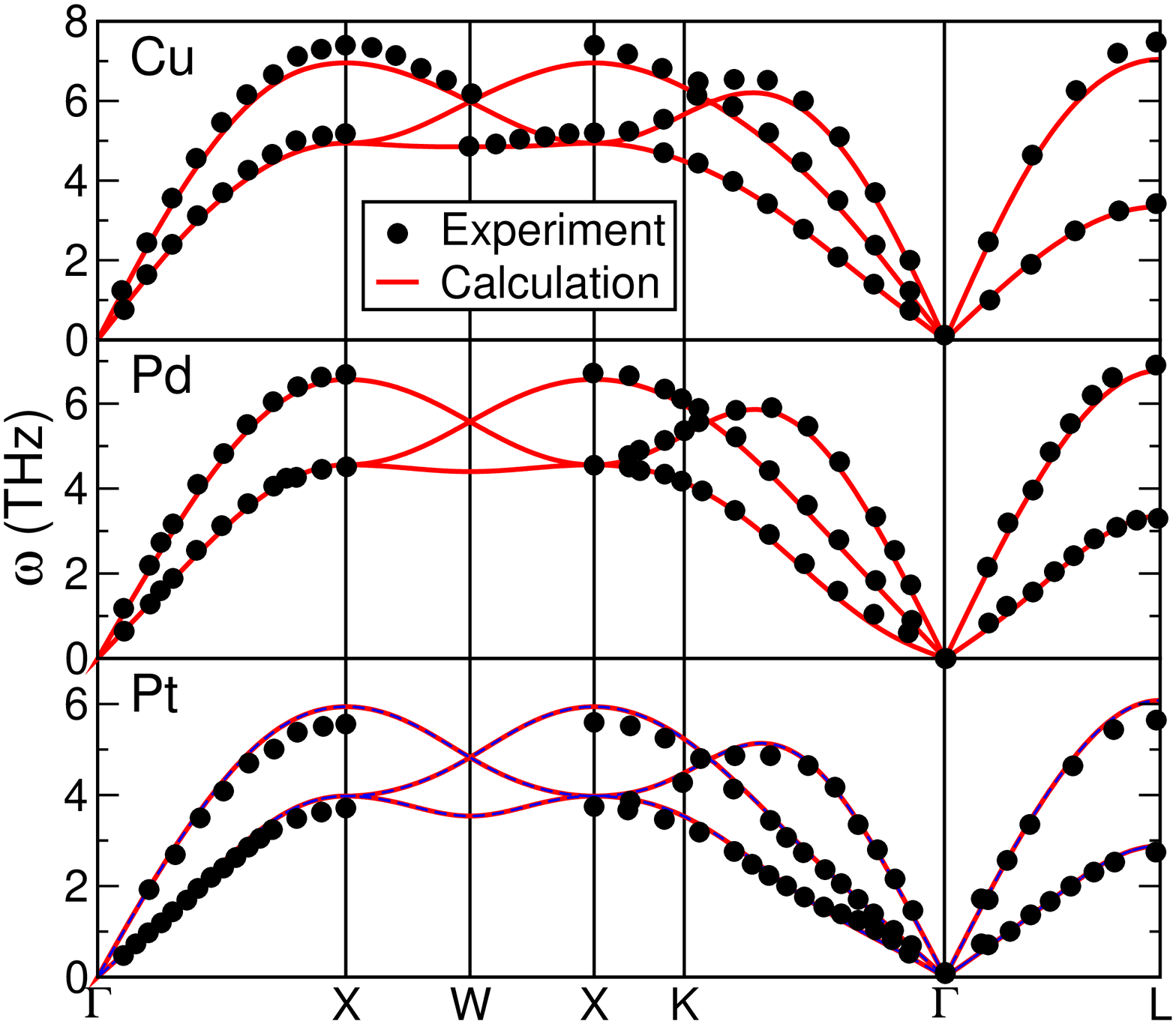}
\caption{Calculated phonon spectra for Cu, Pd and Pt along high-symmetry directions in the fcc Brillouin zone. The experimental data (black dots) are shown for comparison \cite{Schober:81s}. The calculated phonon dispersions of Pt with (solid lines) and without SOC (dashed lines) are nearly the same indicating that SOC has very little effect on phonon modes.}
\label{fig:s1}
\end{figure}

In practice, we use the \textsc{quantum espresso} density functional theory code \cite{Giannozzi:jpcm09s} based on plane-waves and pseudopotentials in combination with the Perdew-Burke-Ernzerhof form for the generalized gradient approximation to the exchange-correlation energy \cite{Perdew:prl96s}. The experimental lattice constants were used for all  metals. The calculated phonon dispersions for Cu, Pd and Pt are shown in Fig.~\ref{fig:s1}. The measured phonon spectra \cite{Schober:81s} are included for comparison and good agreement between experiment and calculation is found in all three cases. The Cu and Pd phonons were calculated without spin-orbit coupling (SOC). For Pt, we calculated the phonons with (red solid lines) and without (blue dashed lines) SOC. Because the interatomic forces are mainly determined by the Coulomb interaction between electrons and nuclei \cite{Baroni:rmp01s} and the main features of the electronic energy bands are not changed by including SOC, the results (are seen in the figure to) lie on top of one another.

We checked that the density functional perturbation theory \cite{Baroni:prl87s} yields the same phonon modes as the force constant approach. The very good agreement between the calculated and measured phonon dispersions seen in Fig.~\ref{fig:s1} indicates that the harmonic approximation upon which both theoretical methods are based captures the most important physics. Should it be necessary to generate correlated lattice disorder including anharmonic effects as input to a transport calculation, first-principles molecular dynamics calculations could be used.

\section{II. Populating phonons and magnons}
Having obtained the phonon energies $\omega_{s\mathbf q}$ and polarization vectors $\bm{\varepsilon}_{s\mathbf q}$ by diagonalizing the dynamical matrix, we are able to populate the phonons in a supercell to generate a configuration of lattice disorder for a chosen temperature. Here $s$ denotes a particular normal mode. At a finite temperature $T$, the vibration of an atom $l$ about its equilibrium position $\mathbf R_l$ can be described by a linear superposition of all occupied normal modes,
\begin{eqnarray}
\!\!\!\!\!\!\!\!\!\! \mathbf u_l(T,t) = \frac{1}{\sqrt{N_qM_l}}\sum_{s\mathbf q}\bm{\varepsilon_{s\mathbf q}}A_{s\mathbf q}(T)e^{i(\mathbf q\cdot\mathbf R_l-\omega_{s\mathbf q}t+\phi_{s\mathbf q})},\label{eq:displace}
\end{eqnarray}
where $N_q$ is the number of wavevectors $\mathbf q$ compatible with the lateral supercell used for the scattering region and $M_l$ is the mass of atom $l$. $\phi_{s\mathbf q}$ is a random phase of the normal mode $\omega_{s\mathbf q}$; varying it allows us to generate different configurations of thermal disorder. In our frozen thermal disorder picture, we can set the time $t$ to be zero without loss of generality. The vibrational amplitude $A_{s\mathbf q}(T)$ is determined by  occupying the phonon mode at the temperature $T$ according to quantum statistics. Specifically, the quantity $\omega^2_{s\mathbf q}A^2_{s\mathbf q}/2$ should equal the total energy contributed by the $s\mathbf q$ phonon mode. 

Magnon modes for bulk Fe are calculated using the ``frozen magnon method'' introduced by Halilov {\it et al.} \cite{Halilov:epl97s,*Halilov:prb98bs} and populated as a function of temperature to generate snapshots of correlated spin disorder by analogy with the phonon case. At a temperature far below the Curie temperature, the occupation of a magnon mode $\mathbf q$ results in a small polar angle $\theta_{\mathbf q}$ of local magnetic moments with respect to the global quantization axis, i.e.,
\begin{equation}
\frac{M_s}{2g\mu_B}\langle\theta^2_{\mathbf q}\rangle=\frac{n_{\mathbf q}(T)}{N_q}.\label{eq:theta}
\end{equation} 
Here $M_s$ is the saturation magnetization, $g$ is the Land{\'e} $g$ factor for the electron, taken to be 2, $\mu_B$ is the Bohr magneton, $\langle\rangle$ denotes thermal averaging and $N_q$ is the total number of magnon modes. The temperature-dependent occupation of the magnon mode $n_{\mathbf q}(T)$ follows Bose-Einstein statistics. The final polar angle of the magnetic moment on every atom results from the linear superposition of $\theta_{\mathbf q}$ for all contributing magnon modes. 

\section{III. Numerical details}
The Kohn-Sham potentials in the atomic spheres approximation (ASA) are calculated self-consistently without SOC using the tight-binding linear muffin-tin orbital (TB-LMTO) method \cite{Andersen:prl84s,*Andersen:prb86s}. Experimental lattice constants are used throughout. For the slab of collinear Ni$_{80}$Fe$_{20}$ binary alloy sandwiched between Cu leads, ASA potentials for Ni and Fe are calculated without SOC using the coherent potential approximation \cite{Soven:pr67s,Turek:97s} combined with a surface Green's function method \cite{Turek:97s} which is also implemented with TB-LMTOs. In the surface Green's function calculations, the two-dimensional Brillouin zone corresponding to an fcc (111) 1$\times$1 interface unit cell is sampled with a 120$\times$120 grid of k points.

SOC makes a negligible contribution to the self-consistent Kohn-Sham potentials and is taken into account adequately in the transport calculation using a Pauli Hamiltonian approach \cite{Daalderop:prb90as}. Such a perturbative treatment has been successfully applied in first-principles calculations of Rashba splitting \cite{Bihlmayer:ss06s}, Dzyaloshinskii-Moriya interaction \cite{Bode:nat07s}, and our own calculations of magnetocrystalline anisotropy \cite{Daalderop:prb90as}, resistivity and magnetization dissipation \cite{Starikov:prl10s,Liu:prl14s,Yuan:prl14s}. For the same reason, the magnon dispersion is calculated without SOC because it is essentially determined by the exchange interactions; the magnetic anisotropy energy of Fe and Ni$_{80}$Fe$_{20}$ is tiny and can be safely neglected. 

\begin{figure}[b]
\includegraphics[width=1\columnwidth]{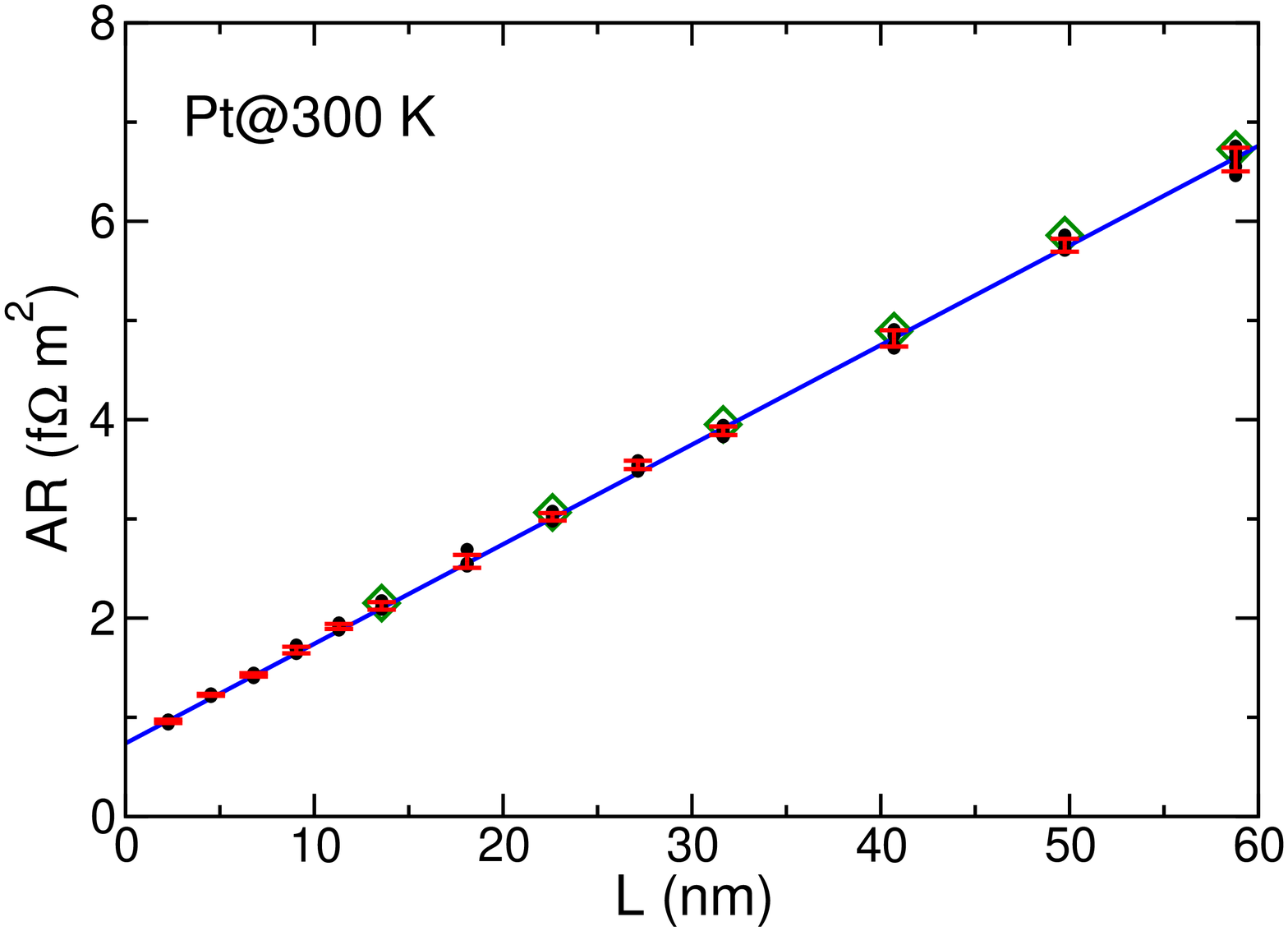}
\caption{Area resistance of Pt calculated as a function of the length of the diffusive Pt (black dots) using a $5\times 5$ lateral supercell. The disordered region of length $L$ is connected to two semi-infinite perfectly crystalline Pt leads and was constructed by populating phonon modes using $T=300$~K. The red bars show the average values and the standard deviation from averaging over more than five random configurations at every length. The solid blue line is the linear least squares fit. The empty green diamonds are resistances calculated by integrating the configuration averaged transmission over the energy window defined by $-\partial f / \partial \varepsilon$ where $f$ is the Fermi-Dirac distribution function with $T=300$~K. The error bars for the diamonds are smaller than the symbol size and hence not shown. }
\label{fig:s2}
\end{figure}
The scattering matrix is determined using a ``wave-function matching'' scheme \cite{Ando:prb91s} also implemented with TB-LMTOs \cite{Xia:prb06s}. For magnetic materials at a finite temperature, the spin-dependent potentials are rotated in spin space \cite{Wang:prb08s} so that the local quantization axis of every atomic sphere conforms to the required spin disorder. The matrix elements of the Pauli Hamiltonian are evaluated using the local quantization axis. We performed numerical tests with lateral supercell sizes up to 10$\times$10 and found that good convergence could be achieved using 5$\times$5 and 4$\times$4 supercells for transport along fcc [111] and bcc [001] directions, respectively. The two-dimensional Brillouin zones of the 5$\times$5 supercell for fcc (4$\times$4 for bcc) metals are sampled with 32$\times$32 (28$\times$28) k points, which are equivalent to 160$\times$160 (112$\times$112) k points in the corresponding 1$\times$1 Brillouin zone.

As a typical example, we plot in Fig.~\ref{fig:s2} the calculated area resistance of Pt as a function of the length of the disordered region. The empty green diamonds show the resistance obtained by configuration averaging the transmission as a function of the energy and then integrating over the energy window defined by the derivative of the Fermi-Dirac distribution function with $T=300$~K (green diamonds) rather than $T=0$~K (red bars). The Fermi smearing has little effect, partly because the lattice disorder already smooths the density of states near the Fermi level so that the conductance varies slowly over an energy range of a few $k_B T$. For this reason, the scattering matrix was only evaluated at the Fermi level in the remainder of this work. 

\begin{figure}[b]
\includegraphics[width=1\columnwidth]{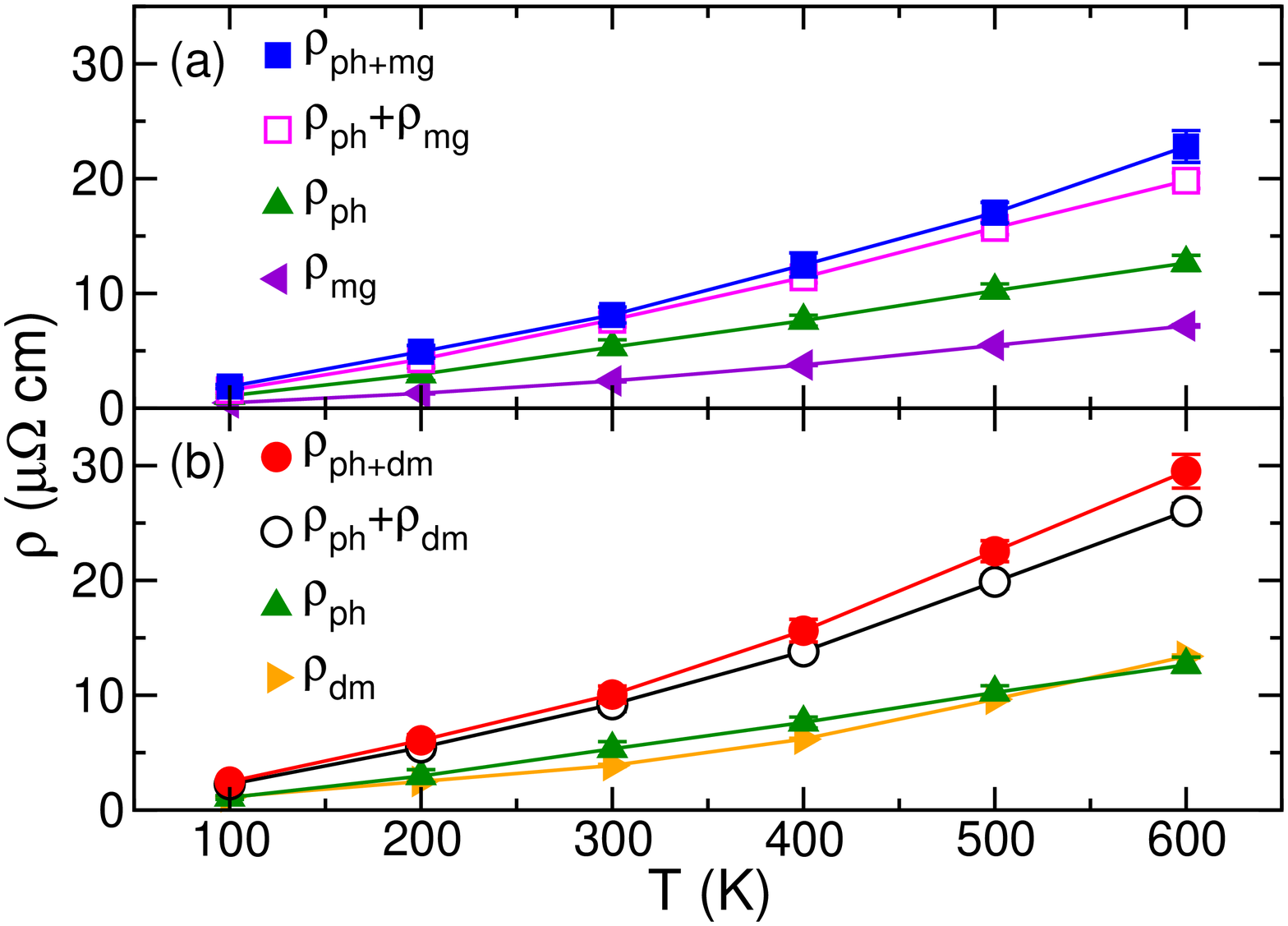}
\caption{Resistivity of Fe calculated with lattice and/or spin disorder. $\rho_{\rm ph}$ (up-pointing, green triangles) is calculated with lattice disorder only, obtained by populating phonon modes while (a) $\rho_{\rm mg}$ (left-pointing, violet triangles) is calculated with spin disorder only, obtained by populating magnon modes. 
$\rho_{\rm ph+mg}$ (solid blue squares), obtained with both lattice and spin disorder simultaneously, is seen to be greater than the sum $\rho_{\rm ph}+\rho_{\rm mg}$ (empty magenta squares).
(b) Instead of calculating spin disorder by populating the magnon spectra, $\rho_{\rm dm}$ (right-pointing, orange triangles) is calculated with spin disorder described using uncorrelated disordered moments. $\rho_{\rm ph+dm}$ (red solid circles) is  calculated with lattice disorder described in terms of phonons and spin disorder described in terms of uncorrelated disordered moments.  $\rho_{\rm ph+dm}$ is greater than the sum $\rho_{\rm ph}+\rho_{\rm dm}$ (empty black circles).
}
\label{fig:s3}
\end{figure}

Fig.~\ref{fig:s2} exhibits Ohmic behavior, i.e. the resistance is proportional to the length of the disordered region and a resistivity value of $\rho=10.0\pm0.3~\mu\Omega$~cm is extracted by a linear least squares fit. The extraction does not depend on the properties of the leads since the Sharvin resistance and other properties of the leads only contribute to the intercept of the linear fit. 
The computing time scales linearly with the length of the scattering region and quadratically with the size of the lateral supercell. Calculating a single configuration of the longest scattering region shown in Fig.~\ref{fig:s2} requires about one hour on a supercomputer node with 32 cores and 256 GB memory; the calculation parallelized perfectly over the two dimensional 32$\times$32 k points summation.

\section{IV. Deviation from Matthiessen's rule}

For ferromagnetic Fe, we can examine the validity of Matthiessen's inequality by comparing the sum of the partial resistivities arising from lattice and spin disorder separately to the total resistivity obtained with both types of disorder present simultaneously. $\rho_{\rm ph}$ and $\rho_{\rm mg}$ in Fig.~\ref{fig:s3}(a) are the resistivities calculated with only phonons and magnons populated in the scattering region, respectively. The sum $\rho_{\rm ph}+\rho_{\rm mg}$ is smaller than the resistivity $\rho_{\rm ph+ mg}$ calculated with both phonons and magnons present simultaneously. The same conclusion can be drawn for the case using uncorrelated spin disorder depicted in Fig.~\ref{fig:s3}(b). Specifically, both $(\rho_{\rm ph}+\rho_{\rm mg})/\rho_{\rm ph+ mg}$ and $(\rho_{\rm ph}+\rho_{\rm dm})/\rho_{\rm ph+ dm}$ are about 0.9 in agreement with a very recent calculation \cite{Glasbrenner:prb14s}.

\end{document}